\documentclass[%
 reprint,
superscriptaddress,
 amsmath,amssymb,
 aps,
 prl,
floatfix,
]{revtex4-1}
\usepackage{multirow}
\usepackage{graphicx}
\usepackage{dcolumn}
\usepackage{bm}
\usepackage{hyperref}
\DeclareUnicodeCharacter{200E}{ }

\begin{document}

\preprint{APS/123-QED}

\title{First observation of the $\beta$3$\alpha$p decay of $^{13}\mathrm{O}$ via $\beta$-delayed charged-particle spectroscopy}
		\author{J.~Bishop}
		\affiliation{Cyclotron Institute, Texas A\&M University, College Station, TX 77843, USA}
		\author{G.V.~Rogachev}
		\affiliation{Cyclotron Institute, Texas A\&M University, College Station, TX 77843, USA}
		\affiliation{Department of Physics \& Astronomy, Texas A\&M University, College Station, TX 77843, USA}
		\affiliation{Nuclear Solutions Institute, Texas A\&M University, College Station, TX 77843, USA}
		\author{S.~Ahn}
		\affiliation{Center for Exotic Nuclear Studies, Institute for Basic Science, 34126 Daejeon, Republic of Korea}
		\author{M.~Barbui}
		\affiliation{Cyclotron Institute, Texas A\&M University, College Station, TX 77843, USA}
		\author{S.M.~Cha}
		\affiliation{Center for Exotic Nuclear Studies, Institute for Basic Science, 34126 Daejeon, Republic of Korea}
		\author{E.~Harris}
		\affiliation{Cyclotron Institute, Texas A\&M University, College Station, TX 77843, USA}
		\affiliation{Department of Physics \& Astronomy, Texas A\&M University, College Station, TX 77843, USA}
		\author{C.~Hunt}
		\affiliation{Cyclotron Institute, Texas A\&M University, College Station, TX 77843, USA}
		\affiliation{Department of Physics \& Astronomy, Texas A\&M University, College Station, TX 77843, USA}
		\author{C.H.~Kim}
		\affiliation{Department of Physics, Sungkyunkwan University (SKKU), Republic of Korea}
		\author{D.~Kim}
		\affiliation{Center for Exotic Nuclear Studies, Institute for Basic Science, 34126 Daejeon, Republic of Korea}
		\author{S.H.~Kim}
		\affiliation{Department of Physics, Sungkyunkwan University, Suwon 16419, Republic of Korea}
		\author{E.~Koshchiy}
		\affiliation{Cyclotron Institute, Texas A\&M University, College Station, TX 77843, USA}
		\author{Z.~Luo}
		\affiliation{Cyclotron Institute, Texas A\&M University, College Station, TX 77843, USA}
		\affiliation{Department of Physics \& Astronomy, Texas A\&M University, College Station, TX 77843, USA}
		\author{C.~Park}
		\affiliation{Center for Exotic Nuclear Studies, Institute for Basic Science, 34126 Daejeon, Republic of Korea}	
		\author{C.E.~Parker}
		\affiliation{Cyclotron Institute, Texas A\&M University, College Station, TX 77843, USA}		
		\author{E.C.~Pollacco}
		\affiliation{IRFU, CEA, Universit\'e Paris-Saclay, Gif-Sur-Yvette, France}		
		\author{B.T.~Roeder}
		\affiliation{Cyclotron Institute, Texas A\&M University, College Station, TX 77843, USA}
		\author{M.~Roosa}
		\affiliation{Cyclotron Institute, Texas A\&M University, College Station, TX 77843, USA}
		\affiliation{Department of Physics \& Astronomy, Texas A\&M University, College Station, TX 77843, USA}		
		\author{A. Saastamoinen}
		\affiliation{Cyclotron Institute, Texas A\&M University, College Station, TX 77843, USA}
		\author{D.P.~Scriven}
		\affiliation{Cyclotron Institute, Texas A\&M University, College Station, TX 77843, USA}
		\affiliation{Department of Physics \& Astronomy, Texas A\&M University, College Station, TX 77843, USA}
	
\email{jackbishop@tamu.edu}

\date{\today}

\begin{abstract}
\begin{description}
\item[Background]
The $\beta$-delayed proton-decay of $^{13}\mathrm{O}$ has previously been studied, but the direct observation of $\beta$-delayed 3$\alpha$p decay has not been reported.
\item[Purpose]
Rare 3$\alpha$p events from the decay of excited states in $^{13}\mathrm{N}^{\star}$ provide a sensitive probe of cluster configurations in $^{13}$N.
\item[Method]
To measure the low-energy products following $\beta$-delayed 3$\alpha$p-decay, the TexAT Time Projection Chamber was employed using the one-at-a-time $\beta$-delayed charged-particle spectroscopy technique at the Cyclotron Institute, Texas A\&M University.
\item[Results]
A total of $1.9 \times 10^{5}$ $^{13}\mathrm{O}$ implantations were made inside the TexAT Time Projection Chamber. 149 3$\alpha$p events were observed yielding a $\beta$-delayed 3$\alpha$p branching ratio of 0.078(6)\%.
\item[Conclusion]
Four previously unknown $\alpha$-decaying excited states were observed in $^{13}$N at 11.3 MeV, 12.4 MeV, 13.1 MeV and 13.7 MeV decaying via the 3$\alpha$+p channel.
\end{description}
\end{abstract}

\pacs{Valid PACS appear here}
\maketitle

\section{\label{sec:Introduction}Introduction}

Exotic neutron-deficient nuclei provide an excellent opportunity to explore new decay modes. Large $\beta$-decay Q-values make it possible to populate proton- or $\alpha$-unbound states in daughter nuclei, paving the way for observation of $\beta$-delayed charged-particle emissions. Reviews of advances in $\beta$-delayed charged-particle emission studies can be found in Ref. \cite{Pfutzner,Blank_2008}, where $\beta$-delayed one, two, and three proton decays as well as $\alpha$p/p$\alpha$ decays are discussed. Here we report on a new decay mode that has not been observed before, the $\beta$3$\alpha$p. Not only do we identify these exotic decays of $^{13}$O, but we were also able to use it to obtain information on cluster structure in excited states of the daughter nucleus, $^{13}$N.

Clustering phenomena are prevalent in light nuclei and are an excellent testing ground for understanding few-body systems that are theoretically accessible. These clustering phenomena have been well-studied in $\alpha$-conjugate nuclei. Much less experimental information is available for N$\ne$Z nuclei. Yet,  theoretical studies (e.g. \cite{Seya,Oertzen1,Kanada}) indicate that cluster configurations may be even richer in non-self-conjugate nuclei, opening a window of opportunity to confront the highly-non-trivial theoretical predictions with experimental data. Recent experimental studies of clustering in non-self-conjugate nuclei already produced exciting results, such as hints for linear chain structures stabilized by ``extra" nucleons (e.g. \cite{Oertzen,Milin,Yamaguchi}) and indications for super-radiance \cite{Marina,Volya}. 

Of particular interest is the nucleus $^{13}\mathrm{N}$ where three $\alpha$ particles and an ``extra'' proton can form exotic cluster configurations. Resonant $^{9}\mathrm{B}$+$\alpha$ scattering or $\alpha$-transfer reactions are not possible because $^{9}\mathrm{B}$ is proton unbound with a half life of the order of $10^{-18}$ s. Instead, one may use $\beta$-delayed charged-particle spectroscopy to populate states in $^{13}\mathrm{N}$ via $^{13}\mathrm{O}$ and observe the decays to a final state of 3$\alpha$p. The $\beta$-delayed proton channel has previously been studied for $^{13}\mathrm{O}$ \cite{Knudsen} where limited statistics showed only a very small sensitivity to populating the p+$^{12}\mathrm{C}(0_{2}^{+})$ (Hoyle state) which results in a $3\alpha$+p final state. Utilizing the Texas Active Target (TexAT) Time Projection Chamber to perform one-at-a-time $\beta$-delayed charged-particle spectroscopy, $\alpha$-decays from the near $\alpha$-threshold excited states in $^{13}$N have been observed for the first time, providing insights into the  $\alpha$+$^{9}\mathrm{B}$ clustering. Capitalizing on the advantages of TPCs for $\beta$-delayed charged-particle emission studies, unambiguous and background-free identifications of the $\beta$3$\alpha$p events were made. Reconstruction of complete kinematics for these exotic decays allowed for robust decay channel assignments, providing insights into the cluster structure of the $^{13}$N excited states. Evidence for the $\frac{1}{2}^{+}$ first excited state in $^{9}\mathrm{B}$, mirror of the well-known $\frac{1}{2}^{+}$ in $^9$Be, was an unexpected byproduct of these measurements, demonstrating the sensitivity of the technique.\par

\section{\label{sec:setup}Experimental setup}
The $\beta$-delayed charged-particle spectroscopy technique with the TexAT TPC has previously been applied for $\beta$-delayed 3$\alpha$ decay studies of $^{12}\mathrm{N}$ via $^{12}\mathrm{C}^{\star}$ \cite{Hoyle}. A detailed description of the technique is provided in \cite{NIM}. Here, we utilize the same experimental approach to observe the $\beta$-delayed 3$\alpha$p decays of $^{13}\mathrm{O}$ via $^{13}\mathrm{N}^{\star}$. We implant $\beta$-decaying $^{13}$O (t$_{1/2}$ = 8.58 ms) one-at-a-time into the TexAT TPC by providing a phase shift signal to the K500 Cyclotron at Texas A\&M University when a successful implantation has taken place to halt the primary beam. This phase shift then lasts for three half-lives or until the observation of a $\beta$-delayed charged particle in TexAT, with the DAQ ready to accept the trigger. The phase shift is then reset to allow for the next implantation. A beam of $^{13}\mathrm{O}$ was produced via the $^3$He($^{14}$N,$^{13}$O) reaction at the MARS (Momentum Achromat Recoil Separator) \cite{MARS} with a typical intensity of 5 pps with an energy of 15.1 MeV/u, degraded by an aluminum foil to 2 MeV/u, to stop inside of the TexAT sensitive area, filled with 50 Torr of CO$_2$ gas. To measure the correlated implantation/decay events, the 2p trigger mode of GET electronics \cite{GET} was employed where the occurrence of two triggers within a 30 ms time window was required for a full event. The first trigger, the L1A (implantation), is generated if the Micromegas pad multiplicity exceeds 10. If, during the 30 ms following the L1A trigger, another trigger occurs with Micromegas pad multiplicity above two, the second L1B (decay) trigger event and the time between the L1A and L1B are recorded. For normalization and beam characterization, all events were recorded, even if L1B trigger never came.

\section{Analysis}
The complete L1A (implant) + L1B (decay) events were selected with the time between the two triggers in the range of 1-30 ms. The short times ($<$1 ms) were omitted to remove double trigger events due to sudden beam-induced noise. To ensure the implanted ion is $^{13}\mathrm{O}$, the energy deposited by the beam implant event in the Micromegas ``Jr" (MM Jr) beam tracker \cite{MMJr} at the entrance to the TexAT chamber was recorded. The beam contaminants were $^{7}\mathrm{Be}$ and $^{10}\mathrm{C}$, dominated by $^{7}\mathrm{Be}$ at $\approx$ 28\% of the beam intensity.\par
Following an identification of $^{13}\mathrm{O}$ implant, the stopping position was evaluated event-by-event using implant tracks, selecting only those which stopped inside the active area of the Micromegas and not closer than 31.5 mm from the edge. The spread of the $^{13}\mathrm{O}$ stopping position inside TexAT was 67.5 mm due to straggling.\par
Further selection was performed by imposing tight correlation ($<$5 mm) between the $^{13}$O stopping location and the vertex location of the respective decay event. Events which passed this test were then fit with a single track segment using a randomly-sampled $\chi$-squared minimization algorithm. If a good fit is achieved, these events were identified as single proton events. The $\beta$-delayed proton spectrum replicates the previous results \cite{Knudsen} well, albeit with decreased resolution that will be covered in a subsequent publication with further experimental details. The remaining events were fit with four track segments as candidates for $\beta$3$\alpha$p decay using randomly-sampled $\chi$-squared minimization. They were then inspected visually to evaluate the fits' quality. Given the complexity of the fits, manual modifications of the fit algorithm parameters were required for some events.

\section{3$\alpha$+proton events\label{sec:3ap}}
\begin{figure}
\centerline{\includegraphics[width=0.5\textwidth]{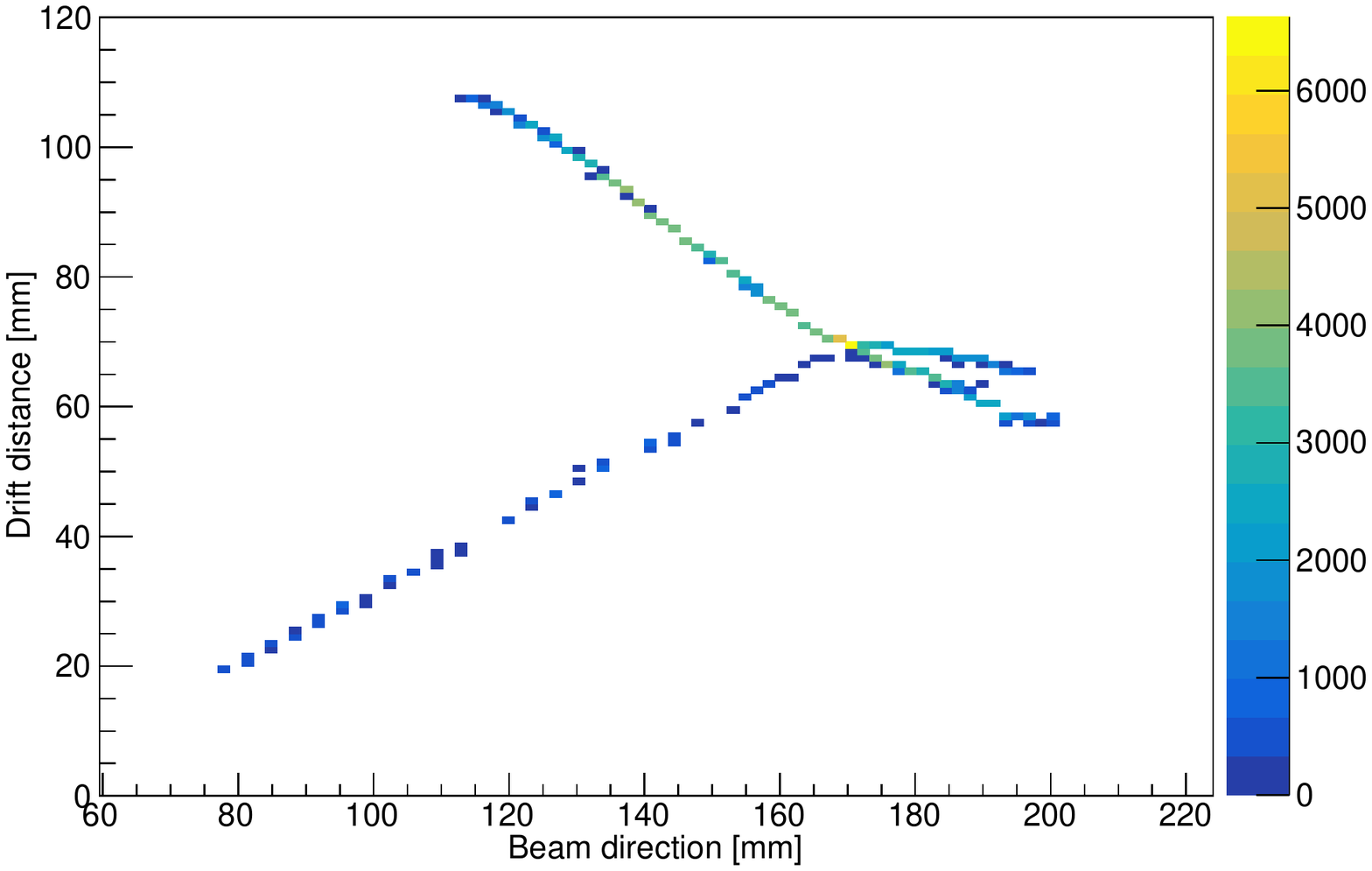}}
\caption{Example 3$\alpha$+p event where the color (online) corresponds to the energy deposition within each voxel after projection into 2D. The proton tracks extends from the vertex to the lower-left of the figure as evidenced by the lower energy deposition. Invariant mass reconstruction designated this event as decaying through the $^{9}$B(g.s)+$\alpha$ channel. \label{fig:event}}
\end{figure}
Overall, 149 $\beta$3$\alpha$p events were identified, an example of which is shown in Fig.~\ref{fig:event}. Due to the size of the TPC and limitations on reconstruction in parts of the TexAT TPC, only 102 out of 149 of these events allow for complete reconstruction. The ``incomplete'' events are dominated by the $^9$B(g.s.)+$\alpha$ decay as this produces a high-energy $\alpha$-particle that may escape from the active volume of the TexAT TPC. The efficiency for the $\alpha_0$ decay starts to deviate from 100\% at $E_{x}$ = 10 MeV, slowly drops to around 60\% at $E_{x}$ = 14 MeV (where $\alpha_i$ signifies $\alpha+^{9}$B decay with $^{9}$B in the i$^{th}$ excited state). The efficiency for $\alpha_1$ and $\alpha_3$ are less affected and only decrease to 70\% at $E_{x}$ = 14 MeV. In proton decays to the Hoyle state, most of the energy is taken by protons and the resulting three $\alpha$-tracks of the pre-selected events are always confined to the active volume of the TPC. Proton tracks were not required in reconstruction as complete kinematics can be recovered from the remaining three $\alpha$-tracks. Therefore, there was no efficiency reduction for the p+$^{12}$C(Hoyle) decays. 

\begin{figure}
\centerline{\includegraphics[width=0.5\textwidth]{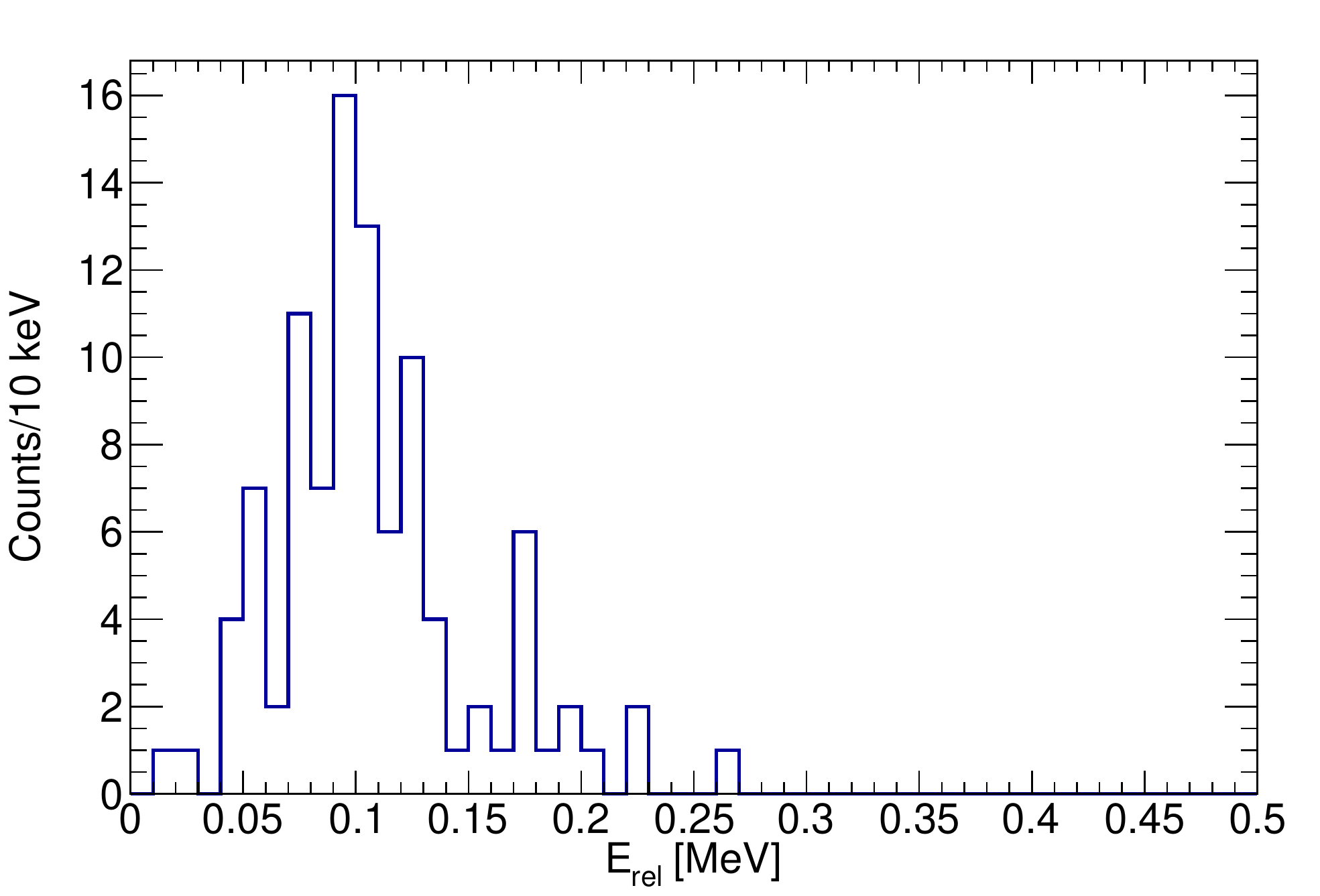}}
\caption{Relative energy spectrum for pairs of $\alpha$-particles with the smallest relative energy of the three $\alpha$-tracks. The $^{8}\mathrm{Be}$(g.s) at 92 keV is well-reproduced.\label{fig:8Be}}
\end{figure}

\begin{figure}
\centerline{\includegraphics[width=0.5\textwidth]{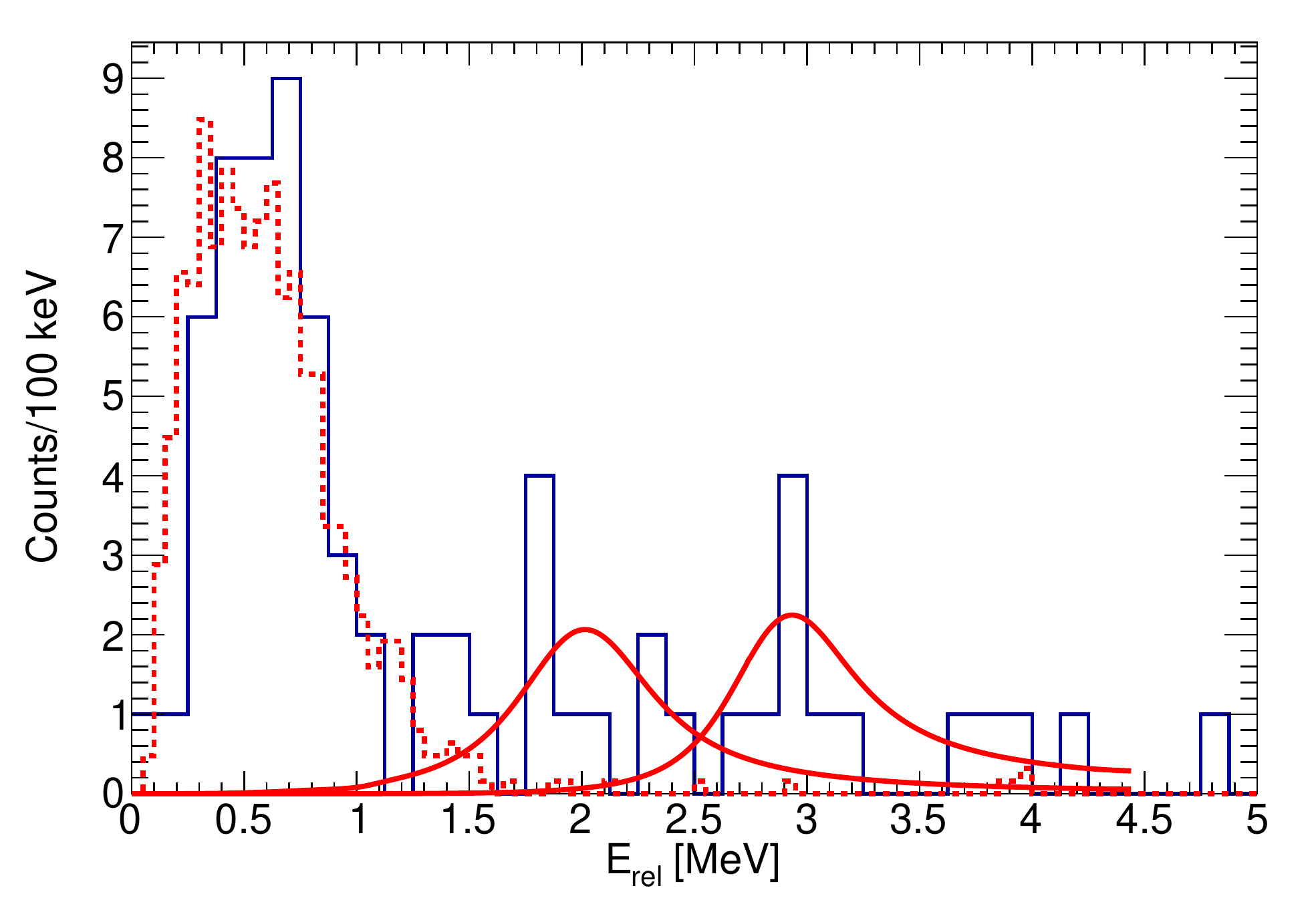}}
\caption{For events that do not decay via the Hoyle state, the relative energy spectrum is shown here which is generated by selecting the two $\alpha$-particles that produce the $^{8}\mathrm{Be}$(g.s) and then reconstructing the $^{9}\mathrm{B}$ relative energy with the proton. Overlaid in dashed red are simulated data for the ground state contribution and in solid red are the $\frac{1}{2}^{+}$ and $\frac{5}{2}^{+}$ states from single channel R-Matrix calculations convoluted with a Gaussian with $\sigma$ = 0.23 MeV. The $\frac{1}{2}^{+}$ parameters are those obtained by Wheldon \cite{Wheldon} which show excellent agreement.\label{fig:9B}}
\end{figure}
\begin{figure}
\centerline{\includegraphics[width=0.5\textwidth]{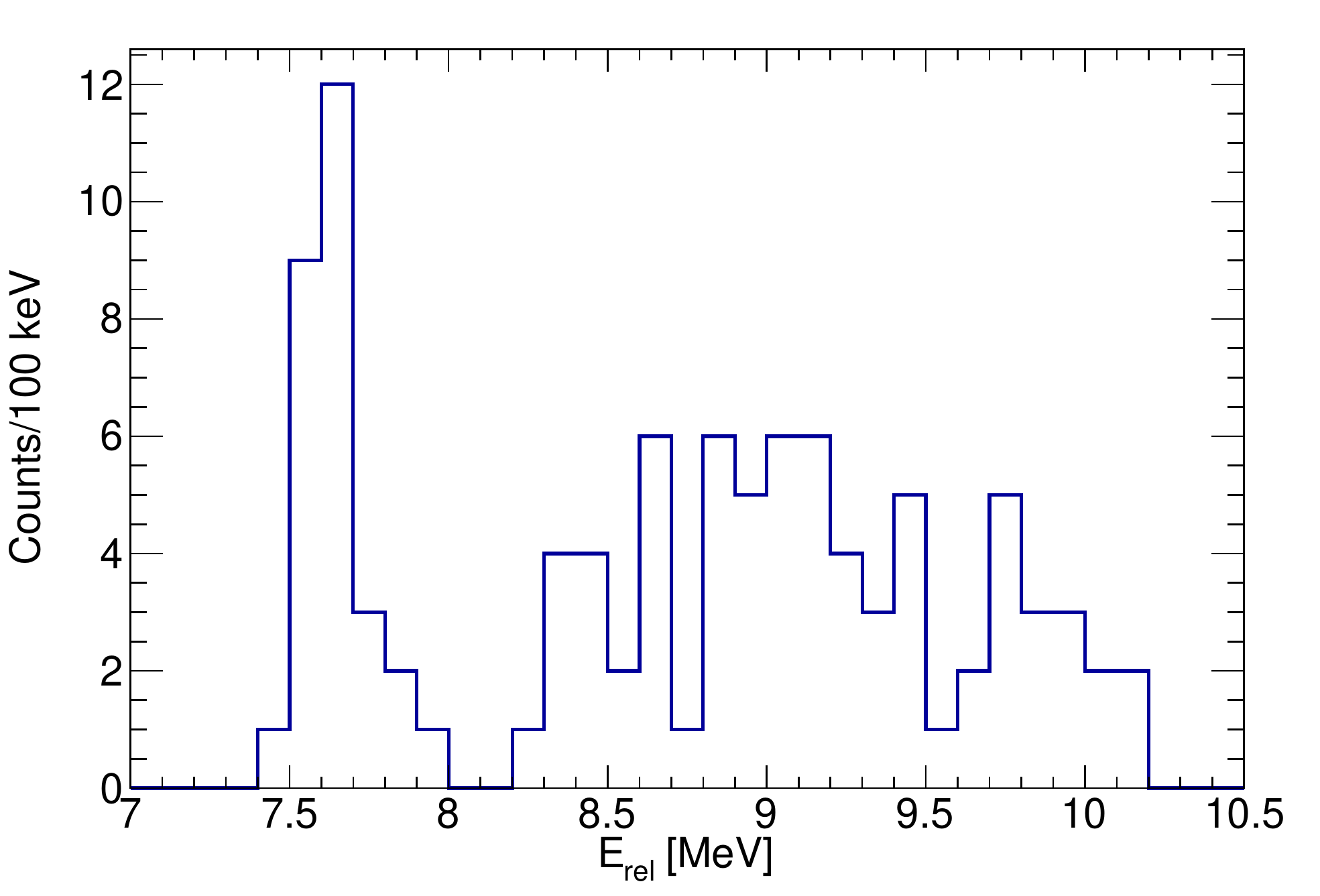}}
\caption{Invariant mass spectrum from 3$\alpha$-particles assuming a $^{12}\mathrm{C}$ origin. A peak at 7.65 MeV is seen, well reproducing the Hoyle state energy and a broad peak is seen at higher excitation energies which correspond to events that decay via $^{9}\mathrm{B}+\alpha$. No peaks from higher excited states in $^{12}$C can be seen.\label{fig:12C}}
\end{figure}
In order to identify the parent state in $^{13}\mathrm{N}^{\star}$, the lowest energy deposition arm was identified as the proton track and the momentum of the 3 $\alpha$-particles was determined by the length and direction of $\alpha$-tracks in the gas. Protons almost always escape the sensitive volume, and the proton momentum is reconstructed from momentum conservation. The decay energy is then the sum of the three $\alpha$-particles' and proton energy. From here, the $^{8}\mathrm{Be}$ (Fig.~\ref{fig:8Be}), $^{9}\mathrm{B}$ (Fig.~\ref{fig:9B}) and $^{12}\mathrm{C}$ (Fig.~\ref{fig:12C}) excitation energies were determined from the invariant mass. This allowed for a selection of events which proceeded to decay via p+$^{12}\mathrm{C}(0_{2}^{+})$ [$p_{2}$], $\alpha$+$^{9}\mathrm{B}$(g.s) [$\alpha_{0}$], $\alpha$+$^{9}\mathrm{B}(\frac{1}{2}^{+})$ [$\alpha_{1}$] and $\alpha$+$^{9}\mathrm{B}(\frac{5}{2}^{+})$ [$\alpha_{3}$]. There is evidence of strength in $^{9}\mathrm{B}$ between 1 and 2.4 MeV excitation energy (Fig.~\ref{fig:9B}). It is likely due to the $\frac{1}{2}^{+}$ state in $^{9}\mathrm{B}$ \cite{Wheldon} that is the mirror of the well-known $\frac{1}{2}^{+}$ first excited state in $^9$Be. Attempts to fit the spectrum without the $\frac{1}{2}^{+}$ in $^9$B fail because it is difficult to explain the excess of counts at excitation energies between 1.4 and 2.4 MeV comparable to the 2.4 - 3.5 MeV region where there are known excited state in $^9$B states. Contributions from a broad $\frac{1}{2}^{-}$ state at 2.78 MeV may give a signature similar to that seen albeit at lower energies (peaking at $E_{rel}$ = 1.3 MeV for a $^{13}\mathrm{N}$($E_{x}$) = 12.4 MeV) when considering the expected yield from a $\frac{1}{2}^{-}$ state in $^{13}\mathrm{N}$. The L=0 $\alpha$-decay to the broad $\frac{1}{2}^{-}$ in $^{9}\mathrm{B}$ will increase the yield at small excitation energies. While this possibility is disfavored from the observed spectrum due to the energy offset, it is mentioned here for completeness. The $\frac{1}{2}^{+}$ state in $^{9}\mathrm{B}$ was selected by taking an excitation energy of between 1.4 and 2.4 MeV in $^{9}\mathrm{B}$ (following the centroid and width as observed via $^{9}\mathrm{Be}(^{3}\mathrm{He},t)$ \cite{Wheldon} which is consistent with our current results) and the $\frac{5}{2}^{+}$ was taken as having an excitation energy of above 2.4 MeV. Any contribution from the relatively-narrow 2.345 MeV $\frac{5}{2}^{-}$ ($\alpha_2$) is not present in the presented plots as this state decays almost exclusively via $^{5}\mathrm{Li}$ and therefore would not correspond to a peak in the $^{8}$Be spectrum. There were only 3 events associated with this decay to $^{5}\mathrm{Li}$ hence the statistics were insufficient to incorporate into the analysis. 

\par
\begin{figure}
\centerline{\includegraphics[width=0.5\textwidth]{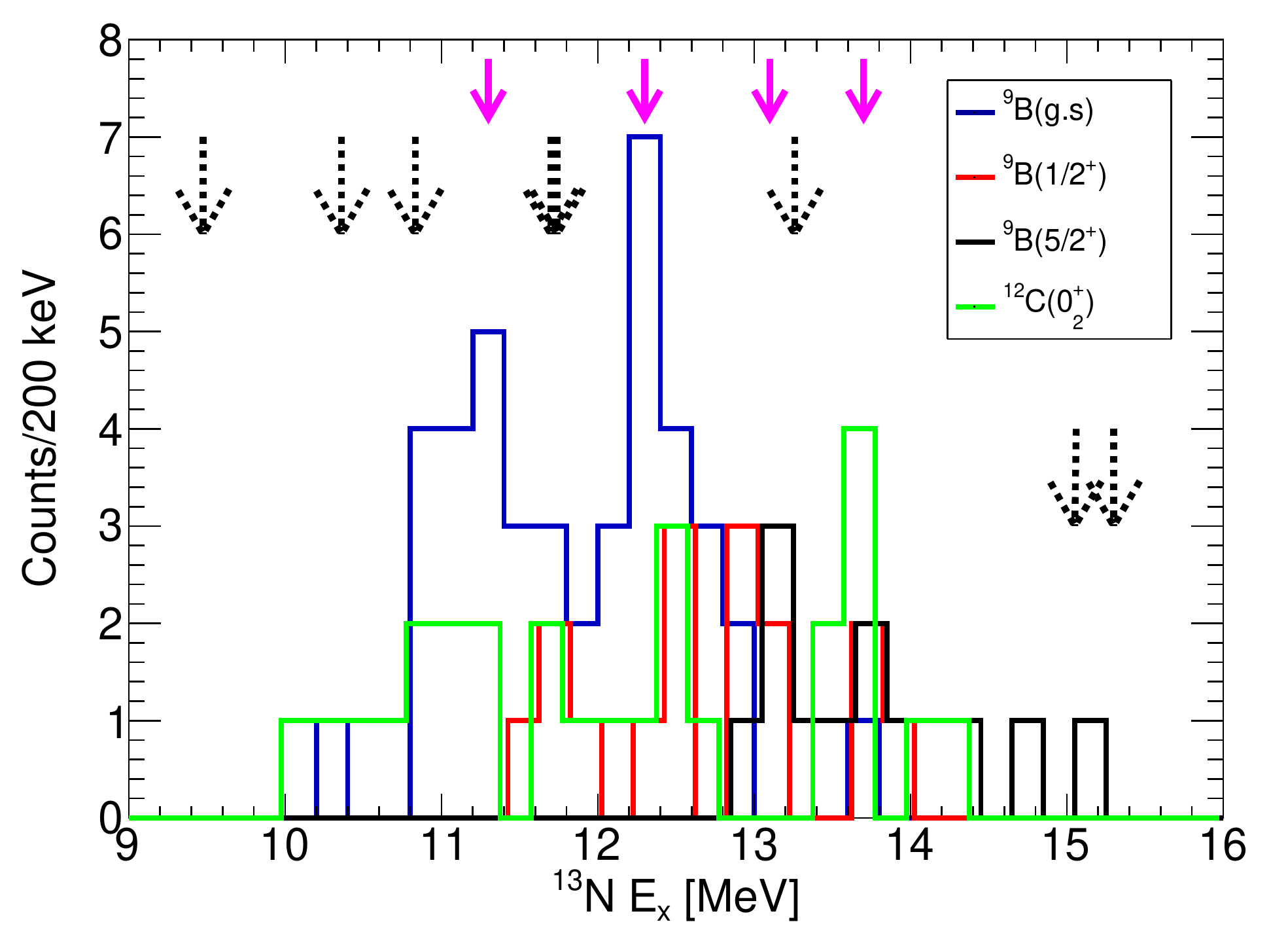}}
\caption{Excitation spectrum in $^{13}\mathrm{N}$ for 3$\alpha+p$ separated by channels. Black dashed arrows show previously-known states populated by $\beta$-decay and new states observed are shown by solid magenta arrows. \label{fig:3ap}}
\end{figure}
\begin{figure}
\centerline{\includegraphics[width=0.5\textwidth]{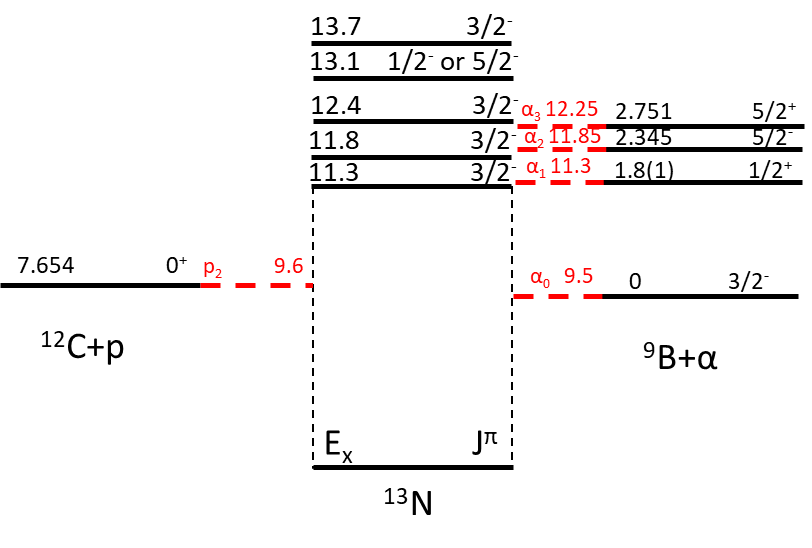}}
\caption{Level scheme of measured 3$\alpha$+p states in $^{13}\mathrm{N}$ in the central column with the proposed spin-parity assignments. The location of the thresholds for proton and $\alpha$ decay are shown in red with the equivalent excitation energy shown. The corresponding states in the daughter nuclei ($^{12}\mathrm{C}$ and $^{9}\mathrm{B}$) are also shown. \label{fig:levelscheme}}
\end{figure}

Following the channel selection, the excitation energy in $^{13}\mathrm{N}$ was calculated and is shown in Fig.~\ref{fig:3ap}. Despite low statistics, a number of states can be seen at 11.3, 12.4, 13.1 and 13.7 MeV. The location of these states relative to the thresholds for $^{9}$B+$\alpha$ and $^{12}\mathrm{C}(0_2^+)$+p is shown in Fig.~\ref{fig:levelscheme}. The clear peak structures (particularly apparent for the $\alpha+^{9}$B(g.s) channel) demonstrate the strength of this technique for studying cluster structures in $^{13}$N. The nuclear structure implications of these states will be the topic of a follow-up paper that also includes more technical detail of the current work. 



\section{Conclusions}
$\beta$-delayed 3$\alpha$p decay has been observed for the first time. While $\beta$-delayed $\alpha$p has been previously observed in $^{9}$C \cite{9C}, $^{17}\mathrm{Ne}$ \cite{17Ne}, $^{21}\mathrm{Mg}$ \cite{21Mg} and $^{23}\mathrm{Si}$ \cite{23Si}, these states did not provide any structural insight and instead were mainly seen through isobaric analogue states that were well fed by $\beta$-decay. In this work, $\beta$3$\alpha$p decay was observed from the states below the isobaric analog in $^{13}\mathrm{N}$ at $E_{x}$ = 15 MeV, demonstrating this is not merely a phase-space effect. The $\beta$-delayed 3$\alpha$p decays observed here are in strong competition with $\beta$-delayed proton decay and therefore the states must have significant clustering. 
Evidence for the low-lying $\frac{1}{2}^{+}$ in $^{9}\mathrm{B}$ in these background-free data, matching the parameters of previous observations \cite{Wheldon}, brings us closer to resolving the long-standing problem of searches for this elusive state. A paper will shortly be published that investigates the properties of the four new states observed here facilitated by this new technique and observed decay channel.
\section{Acknowledgments}
We thank Vlad Goldberg for helpful feedback on this work. This work was supported by the U.S. Department of Energy, Office of Science, Office of Nuclear Science under Award No. DE-FG02-93ER40773 and by the National Nuclear Security Administration through the Center for Excellence in Nuclear Training and University Based
Research (CENTAUR) under Grant No. DE-NA0003841. G.V.R. also acknowledges the support of the Nuclear Solutions Institute. S.A., S.M.C., C.K., D.K., S.K. and C.P. also acknowledge travel support from the IBS grant, funded by the Korean Government under grant number IBS-R031-D1. C.N.K acknowledges travel support from the National Research Foundation of Korea (NRF) grant, funded by the Korea government (MSIT) (No. 2020R1A2C1005981 and 2013M7A1A1075764).
\bibliography{PRCBib}

\end{document}